\pdfoutput=1
\documentclass[copyright,creativecommons]{eptcs}
\providecommand{\event}{ICLP 2026}

\usepackage{amsmath}
\usepackage{amssymb,amsfonts}
\usepackage{mathtools}
\usepackage{bm}
\usepackage{relsize}

\usepackage{xcolor}

\usepackage{graphicx}

\usepackage{xspace}
\usepackage{cleveref}

\usepackage{algorithm}
\usepackage[noend]{algpseudocode}

\usepackage{multicol}
\usepackage{multirow}
\usepackage{verbatimbox}
\usepackage{wrapfig}
\usepackage{fancyvrb}

\usepackage{enumitem}

\usepackage{amsthm}
\theoremstyle{definition}
\newtheorem{definition}{Definition}[section]
\newtheorem{example}[definition]{Example}

\theoremstyle{plain}

\theoremstyle{remark}
\newtheorem{remark}[definition]{Remark}

\usepackage{thmtools,thm-restate}

\newif\ifappendix
\appendixtrue
\ifappendix
  \newcommand{\appref}[2]{Appendix~\ref{#1}}
\else
  \newcommand{\appref}[2]{#2}
\fi

\newcommand{\arxivref}{the full paper~\cite{shapiro2025glp}}

\newenvironment{keywords}{\par\smallskip\noindent\textbf{Keywords:} }{}

\setlist{nosep, leftmargin=*}
\setlist{itemsep=1pt, topsep=3pt, leftmargin=*}

\usepackage{etoolbox}
\usepackage{titlesec}
\makeatletter
\def\thm@space@setup{\thm@preskip=4pt \thm@postskip=4pt}
\makeatother
\BeforeBeginEnvironment{verbatim}{\vspace{-0.3\baselineskip}}
\AfterEndEnvironment{verbatim}{\vspace{-0.4\baselineskip}}
\titlespacing*{\section}{0pt}{8pt}{4pt}
\titlespacing*{\subsection}{0pt}{6pt}{3pt}
\titlespacing*{\subsubsection}{0pt}{4pt}{2pt}

\newcommand{\mypara}[1]{\vspace{1pt}\noindent\textbf{#1.}}
\newcommand{\temph}[1]{\emph{#1}}

\newcommand{\remove}[1]{}

\newcommand{\calV}{\mathcal{V}}
\newcommand{\calG}{\mathcal{G}}
\newcommand{\calT}{\mathcal{T}}
\newcommand{\calF}{\mathcal{F}}

\newcommand{\calS}{\mathcal{S}}

\newcommand{\calC}{\mathcal{C}}

\newcommand{\ia}{\textit{i}}
\newcommand{\ib}{\textit{ii}}

\newcounter{pc}

\providecommand{\titlerunning}{Grassroots Logic Programs}
\providecommand{\authorrunning}{Shapiro}

\providecommand{\thisvolume}[1]{this volume of EPTCS, Open Publishing Association}

\title{GLP: A Grassroots, Multiagent, Concurrent,\\ Logic Programming Language for AI
\ifappendix\texorpdfstring{\\}{ }(Full Version)\fi}

\author{Ehud Shapiro
\institute{London School of Economics and Weizmann Institute of Science}
}

\begin{document}

\maketitle

\begin{abstract}
A grassroots platform is a multiagent distributed system in which multiple independent instances can form and operate independently of each other and of any global resource, yet may coalesce into ever larger instances, possibly resulting in a single global instance.  Grassroots platforms aim to offer an egalitarian/democratic alternative to  centralised/autocratic and decentralised/plutocratic global platforms.

Here, we present Grassroots Logic Programs (GLP), a multiagent concurrent logic programming language designed for the implementation of grassroots platforms:  we recall the standard operational semantics of logic programs; introduce the concurrent operational semantics of   GLP as its restriction; recall multiagent atomic transactions; use them to introduce a multiagent operational semantics of GLP; and prove multiagent GLP to be grassroots. The grassroots social graph---the foundational grassroots platform on which all others are based---serves as a GLP programming example.

These mathematical foundations are being used by AI to implement GLP as well as to program in GLP: a workstation-based implementation of concurrent GLP in Dart was derived from the concurrent operational semantics of GLP; a multiagent smartphone-based implementation of GLP in Dart/Flutter is being developed based on the multiagent operational semantics of GLP; a moded type system for GLP was designed (and implemented by AI in Dart) to facilitate collaborative human-AI development of GLP programs, where AI derives working GLP programs from human-approved type definitions and declarations; GLP implementations of grassroots platforms for the social graph, social networks, currencies and bonds, and more, have been derived by AI from mathematical specifications written as volitional multiagent atomic transactions.

While concurrent logic programming and its powerful programming techniques have been known for four decades, adoption has been hampered by their inaccessibility to the average programmer. With AI as a hyper-programmer, this limitation is removed and the abstract nature and expressive power of concurrent logic programming can be put into effective use.  Our preliminary experience suggests that concurrent logic programming is a more powerful and productive language than mainstream languages for collaborative human-AI abstract specifications-based program development.
\end{abstract}


\section{Introduction}

\mypara{Grassroots} Grassroots platforms aim to offer an egalitarian and democratic alternative to centralised and autocratic (Facebook) and decentralised and plutocratic (Bitcoin) global platforms~\cite{shapiro2025characterising}. A digital platform is \emph{grassroots}~\cite{shapiro2023grassrootsBA,shapiro2025atomic} if it can have multiple instances that can (\ia)~operate independently of each other and of any global resource other than the network, and (\ib)~coalesce into ever larger instances, possibly resulting in a single global instance.  A grassroots platform aims to operate solely on the smartphones of its participants.

Specifications of grassroots platforms that have been provided and proven grassroots include the grassroots social graph~\cite{shapiro2023gsn,shapiro2025atomic} --- the infrastructure platform on which all others build --- grassroots social networks~\cite{shapiro2023gsn,shapiro2026cssn}, grassroots cryptocurrencies and bonds~\cite{shapiro2024gc,lewis2023grassroots,shapiro2026bonds}, and grassroots federations~\cite{shapiro2025GF,halpern2024federated}; the unifying formal framework being volitional multiagent atomic transactions~\cite{lewis2026volitional}. Working GLP implementations of the social graph (this paper), child-safe social networks~\cite{shapiro2026cssn}, grassroots cryptocurrencies and bonds~\cite{shapiro2026bonds}, and secure recovery from major faults~\cite{eitan2026securing} have been derived by AI from these specifications. The Scuttlebutt protocol and social network~\cite{kermarrec2020gossiping} is perhaps the sole example of a deployed grassroots platform to date.

\mypara{Grassroots Logic Programs} Here, we present Grassroots Logic Programs (GLP), a multiagent, concurrent, logic programming language, designed for the implementation of grassroots platforms. Syntactically, GLP extends Logic Programs (LP)~\cite{lloyd1987foundations} with:
\begin{enumerate}
    \item \textbf{Readers:} Each logic variable $X$ (now referred to as \emph{writer}) is paired with a \emph{reader} $X?$, which is assigned a value only once its paired writer $X$ is assigned that value; 
    \item  \textbf{Single-Occurrence (SO):} A variable may occur at most once in a goal or a clause; and
    \item \textbf{Single-Reader Single-Writer (SRSW):} A writer occurs in a clause iff its paired reader does.
\end{enumerate}
The result eschews unification for simple term matching, and conjures both linear logic~\cite{girard1987linear} and futures/promises~\cite{baker1977future,friedman1976impact}: An assignment to a variable may be produced at most once, via the sole occurrence of a writer (promise), and consumed at most once, via the sole occurrence of its paired reader (future).  

Hence, in a multiagent distributed implementation, if a writer and its paired reader are held by different agents, an assignment to the writer is realised as a single message from the writer-holding agent to the reader-holding agent.  The source of all the powerful concurrent logic programming techniques is that such a message may, in turn, contain further readers and writers, enabling the concise expression of rich multidirectional communication modalities.  In the simplest case, a reader sent in a message keeps the channel open for further messages from the sender (streaming); a writer sent in a message allows the receiver to reply.  More generally, variables in a message paired with variables held by third agents can be used for arbitrary network reconfiguration (e.g. friend-mediated introduction in the social graph example below).

\mypara{Semantics} The operational semantics of GLP is presented in two stages: First, we present the \emph{concurrent operational semantics of GLP} (cGLP), an interleaving-based concurrent (single-agent) semantics defined as a restriction of standard LP semantics, preserving computation-as-deduction~\cite{kowalski1974predicate}.

Second, we present multiagent transition systems and their specification via multiagent atomic transactions, and use them to define the \emph{multiagent operational semantics of GLP} (maGLP), with (1) \textbf{Communication:}  Named agents that operate independently while communicating to remote readers assignments made to paired local writers;  (2) \textbf{Cold-calls:} A means for sending a term with variables to a named agent while retaining their paired variables locally.
Cold-calls allow two agents in disconnected components of the social graph to become the owners of paired logic variables. We illustrate maGLP via the grassroots social graph, in which agents may initiate friendships (bidirectional communication channels) via ``cold-calls'' as well as introduce mutual friends to each other.
Lastly, we recall the definition of grassroots platforms and how to prove that a platform specified via atomic transactions is grassroots~\cite{shapiro2023grassrootsBA,shapiro2025atomic}, and apply these to prove that maGLP is indeed grassroots.

\mypara{Historical Context} The concurrent logic programming family emerged in the 1980s --- Concurrent Prolog~\cite{shapiro1983subset}, GHC~\cite{ueda1986guarded}, PARLOG~\cite{clark1986parlog}; surveyed in~\cite{shapiro1989family} --- and was simplified by \emph{flattening} (restricting guards to primitive tests) in Flat Concurrent Prolog (FCP)~\cite{mierowsky1985fcp} and Flat GHC; sequential FCP abstract machines~\cite{houri1989sequential} made commercial deployment feasible. Mode systems followed, including Ueda's moded Flat GHC~\cite{ueda1994moded,ueda1995io,ueda2001resource}, anticipating GLP's SO discipline as a type-system refinement. GLP can be understood as FCP with the SRSW restriction added, simplifying read-only unification~\cite{levi1985readonly}. These languages saw industrial deployment during the Japanese Fifth Generation project~\cite{moto1983overview,shapiro1983fifth}, most prominently in the chat product Virtual Places, developed in FCP by Ubique~\cite{Ubique} and acquired by AOL in 1995, after which the codebase was migrated to C. With the conclusion of the Fifth Generation project~\cite{shapiro19935th}, concurrent logic programming went out of fashion. Its core ideas, however, persisted: pattern-matched message-passing among lightweight share-nothing processes, with single-assignment variables for synchronisation, found their most prominent industrial expression in Erlang~\cite{armstrong2010erlang} --- perhaps the closest descendant of concurrent logic programming --- and from there to Elixir and to actor frameworks such as Akka~\cite{akka2022} in Scala and Orleans~\cite{orleans2022} in .Net. What was lost in this evolution was logic programming's metaprogramming~\cite{safra1988meta,lichtenstein1988concurrent} and the seamless integration of computation and synchronisation through paired logical variables; GLP recovers these, while maintaining the simplicity and efficiency of futures/promises.

\mypara{AI} The foundations presented here are used by AI (Claude) in two distinct disciplines. \emph{Implementing} GLP proceeds through three layers---a mathematical specification, an informal English-and-code specification derived from it by AI, and Dart code derived from that---with authority flowing math$\to$spec$\to$Dart but harmonised by back-and-forth. The concurrent and multiagent operational semantics are first refined into deterministic counterparts, dGLP and madGLP~\cite{shapiro2026implementing}. From dGLP AI derived a workstation-based GLP implementation in Dart, and from madGLP a smartphone-based multiagent one in Dart/Flutter is being developed. Running these implementations surfaced defects, several at the mathematical level, that drove revisions to the operational semantics themselves. A moded type checker was implemented by AI in the same way~\cite{shapiro2026types}. \emph{Programming} in GLP proceeds through three layers of its own---mathematical specifications, provided as guarded multiagent atomic transactions~\cite{lewis2026volitional}; the type definitions and declarations agreed for the program together with the intended behaviour of each procedure; and the GLP code: the designer and AI first agree the types and intent, and only then does AI write, type-check, test, and debug the code, with the human providing design-level oversight rather than code-level intervention~\cite{shapiro2026types}. The type checker catches at compile time a characteristic class of mode errors---confusing a reader for its paired writer---otherwise manifesting as silent run-time suspensions; a befriending clause, for example, was rejected because a reader argument received a produced rather than a consumed value, and AI corrected it from the diagnostic alone. A secure GLP implementation using public-key cryptography, mutual attestations, and friend-based identity custodians is also being developed~\cite{eitan2026securing}. Example GLP programs in this paper are typed.

\mypara{Paper outline}
Section~\ref{sec:glp} presents GLP. Section~\ref{sec:maglp} presents multiagent transition systems, the maGLP definition, and its safety properties. Section~\ref{sec:ma-social-graph} presents the grassroots social graph. Section~\ref{sec:grassroots} proves maGLP is grassroots. Section~\ref{section:conclusion} concludes. \ifappendix All proofs are deferred to the appendices.\else Proofs and supporting material are in  \arxivref.\fi

\section{GLP}
\label{sec:glp}

We present GLP syntax and operational semantics and discuss its programming techniques.

\subsection{Syntax}
\label{sec:glp-ext}

GLP extends LP by adding a paired \emph{reader} $X?$ to every ``ordinary'' logic variable $X$, now called a \emph{writer}.

\begin{definition}[GLP Variables]
\label{def:glp-variables}
Let $\calV$ denote the set of LP variables (identifiers beginning with uppercase), henceforth called \temph{writers}. Define $\calV? = \{X? \mid X \in \calV\}$, called \temph{readers}. The set of all GLP variables is $\hat\calV = \calV \cup \calV?$. A writer $X$ and its reader $X?$ form a \temph{variable pair}.
\end{definition}
GLP terms, unit goals, goals, and clauses are as in LP but defined over the variables in $\hat\calV$.

\begin{definition}[Single-Occurrence (SO) Invariant]
\label{def:so-invariant}
A term, goal, or clause satisfies the \temph{single-occurrence (SO) invariant} if every variable occurs in it at most once.
\end{definition}

\begin{definition}[GLP Program, Goals]
\label{def:glp-program}
A clause $C$ satisfies the \temph{single-reader/single-writer (SRSW) restriction} if it satisfies SO and a variable occurs in $C$ iff its paired variable also occurs in $C$.
A \temph{GLP program} is a finite sequence of clauses satisfying SRSW; clauses for the same predicate form a \temph{procedure}.
The set of GLP goals $\hat\calG(P)$ includes all goals over $\hat\calV$ and the vocabulary of $P$ that satisfy SO.
\end{definition}
The purpose of the SRSW restriction is to prevent multiple writer occurrences racing to assign a variable; the simpler (SW) restriction is insufficient, since if a reader with multiple occurrences is assigned a term with a writer, the result is a writer with multiple occurrences.
We use set notation also when referring to multisets.
\begin{example}[Fair Merge]
\label{ex:merge}
Consider the quintessential concurrent logic program for fairly merging two streams, written in GLP:\footnote{Moded-type definitions (\texttt{T ::= ...}) and declarations (\texttt{procedure ...}) used informally in examples are formally introduced in the Typed GLP companion paper~\cite{shapiro2026types}.}
\begin{verbatim}
Stream ::= [] ;[_|Stream].

procedure merge(Stream?, Stream?, Stream).
merge([X|Xs], Ys, [X?|Zs?]) :- merge(Ys?, Xs?, Zs).
merge(Xs, [Y|Ys], [Y?|Zs?]) :- merge(Xs?, Ys?, Zs).
merge([], Ys, Ys?).
merge(Xs, [], Xs?).
\end{verbatim}
and the goal \verb=merge([1,2,3|Xs?],[a,b|Ys?],Zs)=. The goal satisfies SO and each clause satisfies SRSW. The first clause swaps inputs in the recursive call, ensuring fairness when both streams are available.
\end{example}

\subsection{Operational Semantics}
\label{sec:glp-operational}

\begin{definition}[Transition System, Computation, Run, Safe, Live, Correct~\cite{shapiro2021multiagent,lewis2026volitional}]
\label{def:ts}
A \temph{transition system} is a tuple $TS = (C, c_0, T, {\sim})$ where $C$ is an arbitrary set of \temph{configurations}, $c_0 \in C$ a designated \temph{initial configuration}, $T \subseteq C \times C$ a set of \temph{transitions}, each a pair $c \rightarrow c'$ of non-identical configurations $c \ne c' \in C$, and ${\sim}$ an equivalence relation on $T$; we write $[t]$ for the class of $t \in T$ under ${\sim}$.

A \temph{computation} is a (nonempty, finite or infinite) sequence of configurations $c_1, c_2, \ldots$; it is a \temph{run} if $c_1 = c_0$, and \temph{safe} if $c_i \rightarrow c_{i+1} \in T$ for every two consecutive configurations. We write $c \xrightarrow{*} c'$ for the existence of a safe computation from $c$ to $c'$ (empty if $c = c'$). A class $[t] \in T/{\sim}$ is \temph{enabled} in $c$ if $c \rightarrow c' \in [t]$ for some $c'$, and $c$ is \temph{terminal} if no class is enabled in it. A run is \temph{live} if no class $[t]$ is enabled in every configuration of some suffix in which no member of $[t]$ occurs, and \temph{correct} if it is safe and live. A run is \temph{complete} if it is infinite or ends in a terminal configuration. The \temph{outcome} of a complete run is determined by a domain-specific function from complete runs to an outcome space.
\end{definition}

\begin{definition}[Substitutions and Assignments]
\label{def:writers-assignment}
A GLP \temph{writer assignment} is a term of the form $X := T$, $X\in\calV$, $T\notin\calV$, satisfying SO. Similarly, a GLP \temph{reader assignment} is a term of the form $X? := T$, $X?\in\calV?$, $T\notin\calV$, satisfying SO. A \temph{writers (readers) substitution} $\sigma$ is the substitution implied by a set of writer (reader) assignments that jointly satisfy SO. Given a writers assignment $X := T$, its \temph{readers counterpart} is $X? := T$, and given a writers substitution $\sigma$, its \temph{readers counterpart} $\sigma?$ is the readers substitution defined by $X?\sigma? = X\sigma$. Given a reader assignment $X? := T$, its \temph{writers counterpart} is $X := T$, and given a readers substitution $\tau$, its \temph{writers counterpart} $\tau!$ is the writers substitution defined by $X\tau! = X?\tau$. The \temph{pair completion} of a readers substitution $\sigma$ is $\sigma^\star = \sigma \cup \sigma!$, applied to a fixed point.
\end{definition}

\begin{definition}[GLP Renaming, Renaming Apart]
\label{def:glp-renaming}
A \temph{GLP renaming} is an injective substitution $\rho: \hat\calV \to \hat\calV$ such that for each $X \in \calV$: $X\rho \in \calV$ and $X?\rho = (X\rho)?$. Two GLP terms \temph{have a variable in common} if for some writer $X \in \calV$, either $X$ or $X?$ occurs in both. A GLP renaming $\sigma$ \temph{renames $T'$ apart from} $T$ if $T'\sigma$ and $T$ have no variable in common.
\end{definition}

\begin{definition}[Writer MGU]
\label{def:writer-mgu}
Given two GLP unit goals $A$ and $H$, a \temph{writer mgu} is a writers substitution $\sigma$ such that $A\sigma = H\sigma$ and $\sigma$ is most general among such substitutions. 
\end{definition}

\begin{definition}[GLP Goal/Clause Reduction]
\label{def:glp-reduction}
Given GLP unit goal $A$ and clause $C$, with $H$ \verb|:-| $B$ being the result of the GLP renaming of $C$ apart from $A$, the \temph{GLP reduction} of $A$ with $C$ \temph{succeeds with result} $(B,\sigma)$ if $A$ and $H$ have a writer mgu.
\end{definition}

The cGLP transition equivalence below identifies ``the same transition'' across configurations as goals are instantiated by reader substitutions; following~\cite{lewis2026volitional}, we first assign each goal a persistent identity.

\begin{definition}[Goal Identity]
\label{def:goal-identity}
Each unit goal in a cGLP computation carries a unique \temph{identifier} assigned at spawn time: goals in the initial goal $G_0$ receive distinct initial identifiers, and goals spawned by a Reduce transition receive fresh identifiers. Communicate transitions preserve identifiers: the goal containing $X?$ retains its identity after instantiation.
\end{definition}

\begin{definition}[cGLP Transition System]
\label{def:cglp-ts}
Given a GLP program $P$, an \temph{asynchronous resolvent} over $P$ is a pair $(G, \sigma)$ where $G \in \hat\calG(P)$ and $\sigma$ is a readers substitution.

A transition system $cGLP = (\calC, c_0, \calT, {\sim})$ is a \temph{cGLP transition system} over $P$ and initial goal $G_0$ satisfying SO if:
\begin{enumerate}
    \item $\calC$ is the set of all asynchronous resolvents over $P$
    \item $c_0 = (G_0, \emptyset)$
    \item $\calT$ is the set of all transitions $(G, \sigma) \rightarrow (G', \sigma')$ satisfying either:
    \begin{enumerate}
        \item \textbf{Reduce:} there exists unit goal $A \in G$ such that $C \in P$ is the first clause for which the GLP reduction of $A$ with $C$ succeeds with result $(B, \hat\sigma)$, $G' = (G \setminus \{A\} \cup B)\hat\sigma$, and $\sigma' = \sigma \circ \hat\sigma?$
        \item \textbf{Communicate:} $\{X? := T\} \in \sigma$, $X?\in G$, $G' = G\{X? := T\}$, and $\sigma' = \sigma$
    \end{enumerate}
    \item ${\sim}$, the \temph{cGLP transition equivalence}, relates $t_1 \sim t_2$ iff either both are Reduce transitions reducing the same goal (by identity, Definition~\ref{def:goal-identity}) with the same clause, or both are Communicate transitions applying the counterpart of the same writer assignment to the same goal (by identity)
\end{enumerate}
\end{definition}

cGLP Reduce differs from LP in (1) the use of a writer mgu instead of a regular mgu and (2) the choice of the first applicable clause instead of any clause. The first is the fundamental use of GLP readers for communication and synchronisation. The second compromises on the or-nondeterminism of LP to allow writing fair concurrent programs, such as fair merge above. Note that or-nondeterminism is not completely eliminated, as different scheduling of arrival of assignments on the two input streams of \verb|merge| may result in different orders in its output stream.

The cGLP Communicate rule realises the use of reader/writer pairs for asynchronous communication: it communicates an assignment from its writer to its paired reader.

\mypara{Monotonicity}
In LP, if a goal cannot be reduced, it will never be reduced. In cGLP, a goal that cannot be reduced now may be reduced in the future: if $A$ and $H$ have an mgu that writes on a reader $X? \in A$, and therefore have no writer mgu at present,  another goal that has $X$ may reduce, assigning $X$, and later $X?$, to a value that will allow $A$ and $H$ to have a writer mgu. Conversely, in LP, if a goal $A$ can be reduced now with some clause $H$\verb|:-|$B$, with a regular mgu of $A$ and $H$, it may not be reducible in the future due to variables that $A$ shares with other goals being assigned values by reductions of other goals, preventing unification between the instantiated $A$ and $H$. In cGLP, if a goal $A$ can be reduced now (with a writers mgu), it can always be reduced in the future, as the SO invariant ensures that no other goal can assign any writer in $A$.

Implementation-wise, if a GLP goal $A$ cannot be reduced now, but there is a readers substitution $\sigma$ such that $A\sigma$ can be reduced, such readers are identified, the goal $A$ \emph{suspends} on these readers, and is rescheduled for another reduction attempt once any of them is assigned.

Despite these differences, cGLP has the same notion of logical consequence as LP.

\begin{definition}[cGLP Proper Run, Outcome]
\label{def:cglp-proper-run}
A cGLP run $\rho: (G_0,\sigma_0) \rightarrow \cdots \rightarrow (G_n, \sigma_n)$ is \temph{proper} if for any $1\le i< n$, a variable that occurs in $G_{i+1}$ but not in $G_i$ also does not occur in any $G_j$, $j<i$; the \temph{outcome} of a proper run is $(G_0$ \texttt{:-} $G_n)\sigma_n^\star$; and the run is \temph{successful} if $G_n=\emptyset$.
\end{definition}
The outcome employs pair completion (Definition~\ref{def:writers-assignment}) since $\sigma_n$ accumulates only readers counterparts, leaving the writers of $G_0$ unbound. Let $/?$ be an operator that replaces every reader by its paired writer.

\begin{restatable}[cGLP Computation is Deduction]{proposition}{propGLPDeduction}
\label{prop:cglp-deduction}
Let $(G_0$ \texttt{:-} $G_n)\sigma_n^\star$ be the outcome of a proper cGLP run $\rho: (G_0,\sigma_0) \rightarrow \cdots \rightarrow (G_n, \sigma_n)$. Then $(G_0$ \texttt{:-} $G_n)\sigma_n^\star/?$ is a logical consequence of $P/?$.
\end{restatable}

We note two additional safety properties of cGLP runs.

\begin{restatable}[SO Preservation]{proposition}{propSOPreservation}
\label{prop:so-preservation}
If the initial goal $G_0$ satisfies SO, then every goal in a proper cGLP run satisfies SO.
\end{restatable}

\begin{restatable}[Monotonicity]{proposition}{propMonotonicity}
\label{prop:glp-monotonicity}
In any proper cGLP run, if unit goal $A$ can reduce with clause $C$ at step $i$, then either an instance of $A$ has been reduced by step $j > i$, or an instance of $A$ can still reduce with $C$ at step $j$.
\end{restatable}

Every cGLP run is safe, so correctness (Definition~\ref{def:ts}) reduces to liveness.

\begin{definition}[cGLP Correct Run]
\label{def:cglp-fair-run}
A cGLP run is \temph{correct} iff for every equivalence class $[t] \in \calT/{\sim}$, whenever $[t]$ is enabled some $t' \in [t]$ is eventually taken.
\end{definition}

The SO invariant of GLP allows eschewing unification in favour of \emph{term matching}: if two terms that jointly satisfy SO are unifiable, their mgu maps each variable in one term to a subterm of the other. Term matching thus performs joint term-tree traversal and collects variable assignments along the way; the detailed definition and table appear in \appref{appendix:term-matching}{\arxivref}.

\subsection{Guards}
\label{sec:guards}

GLP clauses may include \emph{guards}---tests that determine clause applicability.

\begin{definition}[Guarded Clause]
\label{def:guarded-clause}
A \temph{guarded clause} has the form $H$ \verb|:-| $G$ \verb"|" $B$, where $H$ is the head, $G$ is a conjunction of guard predicates, and $B$ is the body. The guard separator ``\verb"|"'' distinguishes guards from the body, and is interpreted logically as a conjunction.  Guard arguments are readers paired to head writers.
\end{definition}

Guards have three-valued semantics. Each guard predicate explicitly defines its \emph{success} condition. A guard \emph{suspends} if it does not succeed but some instance of it under a readers substitution would succeed. A guard \emph{fails} if no such instance exists. A guard conjunction succeeds if all members succeed; it suspends if any member suspends and none fail; it fails if any member fails.

Definition~\ref{def:glp-reduction} of a GLP goal/clause reduction is augmented to succeed if the guard also succeeds.

\begin{remark}[Guards and SRSW]
\label{rem:guards-srsw}
Guard occurrences count toward SRSW satisfaction: if $X?$ occurs in a guard, its paired writer $X$ must occur in the head and $X?$ may additionally occur once in the body.
Furthermore, if the success of a guard implies that $X?$ is bound to a ground term, then both $X$ and $X?$ may occur multiple times in the clause. Groundness-implying guards include \verb|ground|, \verb|integer|, \verb|number|, \verb|string|, \verb|constant|, arithmetic comparisons (\verb|<|, \verb|>|, \verb|=<|, \verb|>=|, \verb|=:=|, \verb|=\=|), and ground equality (\verb|=?=|). However,  \verb|known| and \verb|compound| do not imply groundness.
\end{remark}

\begin{remark}[Anonymous Variables]
\label{rem:anonymous-variables}
An \emph{anonymous variable} is any variable whose name begins with \verb|_| (e.g., \verb|_|, \verb|_Out|). Anonymous variables may appear anywhere a writer variable may appear; each occurrence denotes a fresh writer with no paired reader. Anonymous readers (\verb|_?|, \verb|_Name?|) are not permitted.
\end{remark}

\mypara{System predicates and body kernels}
GLP \emph{system predicates} such as \verb|:=| (arithmetic assignment), \verb|now| (clock access), and \verb|=..| (term composition/decomposition) are predicates defined by GLP clauses whose bodies may invoke \emph{body kernel predicates}---runtime-implemented primitives not directly accessible to user programs. The system predicate's own guards ensure that body kernel preconditions are met before invocation. For example, arithmetic assignment is defined recursively by clauses that pattern-match on each operator:

\begin{verbatim}
Result? := N :- number(N?) | Result = N?.
Result? := X + Y :- number(X?), number(Y?) |
    '_add'(X?, Y?, Result).
Result? := X + Y :- otherwise |
    X1 := X?, Y1 := Y?, Result := X1? + Y1?.
\end{verbatim}
\noindent The base case binds a plain number; the first addition clause guards that both operands are numbers and invokes the \verb|'_add'| body kernel; the fallback clause recursively evaluates subexpressions. The full table of body kernels and the guard reference appear in \appref{appendix:guards-system}{\arxivref}; the complete system-predicate definitions are in \verb|programs/self.glp| of the repository.

\subsection{GLP Programming Techniques}

On the one hand, GLP has no unification, only pattern-matching and single-writer variables. On the other hand, goal reduction may write on several variables atomically. The result is that almost the entire range of concurrent logic programming techniques developed since the 1980s~\cite{shapiro1989family,shapiro1987concurrent} is available in GLP, including: streams and their fair, biased, and self-balancing merge~\cite{shapiro1984fair,shapiro1986multiway}; channels as generalised streams~\cite{tribble1988channels}; messages with reply variables and bounded buffers~\cite{takeuchi1988bounded}; network reconfiguration and distributed programming~\cite{shafrir1988distributed}; object-oriented programming with mutable state via stream-recursion~\cite{shapiro1983object}; systems programming and computation control~\cite{shapiro1984systems,silverman1988logix}; and the full range of meta-interpreters --- fail-safe, termination-detecting, tracing, and algorithmic debugging~\cite{safra1988meta,lichtenstein1988concurrent}. Tested GLP renditions of these and further techniques --- monitors, distributors, network switches, replicators, and the metainterpreter family --- are in the \verb|programs| directory of the public GLP repository.\footnote{\url{https://github.com/EShapiro2/GLP}}

Two techniques require workarounds. The use of logic variables for mutual exclusion --- standard in concurrent logic languages with atomic unification~\cite{shapiro1989family} --- must be achieved in GLP via a monitor~\cite{hoare1974monitors,hansen1973operating}, the standard distributed-computing solution. Similarly, general broadcast --- distributing a value to multiple consumers --- is not directly supported by GLP, since it would require multiple occurrences of the same reader, in violation of SO; it must be achieved by explicit \emph{distributors}.  We note that if moded-type checking can determine that the values to be distributed never contain writers, then the SO restriction on readers may safely be relaxed allowing broadcast natively; a future upgrade of the language.  Excluded is the encoding of logic programs or-parallel by the and-parallelism of Concurrent Prolog~\cite{codish1986compiling,shapiro1989or}, which depends on general unification in a fundamental way.

\section{Multiagent GLP}\label{sec:maglp}

We extend GLP to multiple agents: Recall the notion of multiagent transition systems via multiagent atomic transactions~\cite{shapiro2021multiagent,shapiro2025atomic};  define multiagent GLP (maGLP) as a transactions-based multiagent transition system; and establish its safety properties. 

\subsection{Multiagent Transition Systems}
\label{sec:mts}
We assume a potentially infinite set of \emph{agents} $\Pi$, but consider only finite subsets of it, so when we refer to a particular set of agents $P \subset \Pi$ we assume $P$ to be nonempty and finite. We use $\subset$ to denote the strict subset relation and $\subseteq$ when equality is also possible.

We use $S^P$ to denote the set $S$ indexed by the set $P$, and if $c\in S^P$ we use $c_p$ to denote the member of $c$ indexed by $p\in P$. Intuitively, think of such a $c\in S^P$ as an array of cells indexed by members of $P$ with cell values in $S$.

\begin{definition}[Local States, Configuration, Transaction, Participants]\label{def:transaction}
Given agents $Q \subset \Pi$ and an arbitrary set $S$ of \temph{local states}, a \temph{configuration} over $Q$ and $S$ is a member of $C:= S^Q$. An \temph{atomic transaction}, or just \emph{transaction}, over $Q$ and $S$ is any pair of configurations $t=c\rightarrow c' \in C^2$ such that $c\ne c'$, with $t_p := c_p \rightarrow c'_p$ for any $p\in Q$, and with $p$ being an \temph{active participant} in $t$ if $c_p\ne c'_p$, \temph{stationary participant} otherwise.
\end{definition}

\begin{definition}[Degree]\label{def:degree}
The \temph{degree} of a transaction $t$ (unary, binary, \ldots, $k$-ary) is the number of active participants in $t$, and the \temph{degree} of a set of transactions $T$ is the maximal degree of any $t\in T$.
\end{definition}

\begin{definition}[Multiagent Transition System]\label{def:mts}
Given agents $P \subset \Pi$ and an arbitrary set $S$ of \temph{local states} with a designated \temph{initial local state} $s_0\in S$, a \temph{multiagent transition system} over $P$ and $S$ is a transition system $TS= (C,c_0,T,{\sim})$ with \temph{configurations} $C:= S^P$, \temph{initial configuration} $c_0:= \{s_0\}^P$, \temph{transitions} $T\subseteq C^2$ being a set of transactions over $P$ and $S$, and ${\sim}$ an equivalence on $T$; the \temph{degree} of $TS$ is the degree of $T$.
\end{definition}

Rather than specifying a multiagent transition system over a set of agents $P$ directly, we specify it via atomic transactions, which are typically of bounded degree smaller than $|P|$.

\begin{definition}[Transaction Closure]\label{def:closure}
Let $P\subset \Pi$, $S$ a set of local states, and $C:=S^P$.
For a transaction $t=(c\rightarrow c')$ over local states $S$ with participants $Q \subseteq P$, the \temph{$P$-closure of $t$}, $t{\uparrow}P$, is the set of transitions over $P$ and $S$ defined by:
$$
t{\uparrow}P := \{ t' \in C^2 :
\forall q\in Q.(t_q = t'_q) \wedge \forall p\in P\setminus Q.(p\text{ is stationary in }t')\}
$$
If $R$ is a set of transactions, each $t\in R$ over some $Q\subseteq P$ and $S$, then the \temph{$P$-closure of $R$}, $R{\uparrow}P$, is the set of transitions over $P$ and $S$ defined by:
$$
R{\uparrow}P := \bigcup_{t\in R} t{\uparrow}P
$$
Given an equivalence ${\sim}$ on $R$, its \temph{$P$-closure} ${\sim}{\uparrow}P$ is the relation on $R{\uparrow}P$ with $\hat t \mathrel{({\sim}{\uparrow}P)} \hat t'$ iff $\hat t \in t{\uparrow}P$ and $\hat t' \in t'{\uparrow}P$ for some $t \sim t'$.
\end{definition}

Namely, the closure over $P\supseteq Q$ of a transaction $t$ over $Q$ includes all transitions $t'$ over $P$ in which members of $Q$ do the same in $t$ and in $t'$, and the rest remain in their current (arbitrary) state. The closure likewise carries any equivalence on transactions to one on the induced transitions: since distinct transactions over the same participants have disjoint closures, ${\sim}{\uparrow}P$ is an equivalence whenever ${\sim}$ is.

\begin{definition}[Transactions-Based Multiagent Transition System]\label{def:tbmts}
Given agents $P \subset \Pi$, local states $S$ with initial local state $s_0\in S$, a set of transactions $R$, each $t\in R$ over some $Q\subseteq P$ and $S$, and an equivalence ${\sim}$ on $R$, the \temph{transactions-based multiagent transition system} over $P$, $S$, $R$, and ${\sim}$ is the multiagent transition system $TS= (S^P,\{s_0\}^P,R{\uparrow}P,{\sim}{\uparrow}P)$.
\end{definition}

\subsection{From cGLP to maGLP}
\label{sec:glp-to-maglp}\label{sec:maglp-def}

In extending GLP to multiple agents, each agent maintains its own asynchronous resolvent as its local state. The key insight is that GLP's variable pairs provide natural binary communication channels: when agent $p$ assigns a writer $X$ for which the paired reader $X?$ is held by agent $q$, the assignment $X := T$ must be communicated to $q$.

A key difference between cGLP and maGLP is in the initial state. In a multiagent transition system all agents must have the same initial local state $s_0$ (Definition~\ref{def:mts}),  precluding the initial sharing of logic variables, as this would imply different initial states for different agents.
We resolve this in two steps. First, we employ only anonymous logic variables ``\verb|_|'' in the initial local states of agents: Anonymous variables are, on the one hand, syntactically identical, hence allow all initial states to be syntactically identical, and on the other hand represent unique variables, hence semantically all initial goals have unique, local, non-shared variables. The initial state of all agents is the atomic goal \verb|agent(ch(_?,_),ch(_?,_))|, providing two bidirectional channels: the first to the person operating the machine, and the second to the network. The first channel is a bidirectional stream of GLP messages between the machine and the person, mediated by runtime UI support below the GLP layer and by physical UI hardware; messages flowing toward the person are rendered as UI elements, and the person's responses flow back as GLP messages. With appropriate UI runtime support, a GLP message to the user may contain a writer, in which case the runtime presents it as a question and an eventual response from the person becomes the corresponding assignment. The second channel carries communication with other agents.

Second, the Cold-call transaction enables agents to bootstrap communication by establishing shared variables through the network infrastructure, realising the cold-call protocol for connecting previously-disconnected agents.

\begin{definition}[Multiagent GLP]\label{definition:maGLP}
Given agents $P\subset \Pi$ and GLP program $M$, the \temph{maGLP transition system} over $P$ and $M$ is the transactions-based multiagent transition system (Definition~\ref{def:tbmts}) over $P$, local states being asynchronous resolvents $(G_p, \sigma_p)$ over $M$, initial local state $s_0 = (\{\verb|agent(ch(_?,_),ch(_?,_))|\}, \emptyset)$, the following transactions $c\rightarrow c'$, and the equivalence ${\sim}$ given below:
\begin{enumerate}
    \item \textbf{Reduce $p$:} A unary transaction with participant $p$ where $c_p \rightarrow c'_p$ is a cGLP Reduce transition (Definition~\ref{def:cglp-ts}).

    \item \textbf{Communicate $p$ to $q$:} A transaction with participants $p,q\in P$ where $\{X? := T\} \in \sigma_p$, $X?$ occurs in $G_q$, $\sigma'_p = \sigma_p \setminus \{X? := T\}$, and $c'_q = (G_q\{X? := T\}, \sigma_q)$.

    \item \textbf{Cold-call $p$ to $q$:} A binary transaction with participants $p\ne q \in P$ where the network output stream in $c_p$ has a new message \verb|msg|$(q,X)$, $c'_p$ is the result of advancing the network output stream in $c_p$, and $c'_q$ is the result of adding \verb|msg|$(q,X)$ to the network input stream in $c_q$.

    \item ${\sim}$, the \temph{maGLP transaction equivalence} on the transactions above, relates $t_1 \sim t_2$ iff both are Reduce transactions at the same agent $p$ reducing the same goal (by identity, Definition~\ref{def:goal-identity}) with the same clause, both are Communicate transactions from the same $p$ to the same $q$ applying the counterpart of the same writer assignment to the same goal (by identity), or both are Cold-call transactions from the same $p$ to the same $q$ delivering the same message.
\end{enumerate}
\end{definition}

Note that Communicate may be unary or binary, depending on whether $p=q$.  Communicate transfers assignments from writers to readers between agents sharing a paired reader and writer.  Cold-call transfers a term with its variables to $q$ through the network streams established in each agent's initial configuration, enabling the creation of paired variables among previously-disconnected agents.
Cold-call is the exception, as once agents share a paired variable they can use it to communicate indefinitely; moreover, an agent with two friends (with which it shares channels) may introduce them to each other, also eschewing the need for a Cold-call.

\subsection{Safety Properties of maGLP}
\label{sec:maglp-safety}

The safety properties established for cGLP in Section~\ref{sec:glp} extend to maGLP. SO preservation (cf.\ Proposition~\ref{prop:so-preservation}) generalises directly:

\begin{restatable}[maGLP SO Preservation]{proposition}{propMaGLPSOPreservation}
\label{prop:maglp-so-preservation}
If the initial goals of all agents satisfy SO, then every goal in every agent's resolvent throughout a proper run satisfies SO.
\end{restatable}

\begin{definition}[maGLP Proper Run, Outcome]
\label{def:maglp-proper-run}
An maGLP run is \temph{proper} if the run at each agent is proper (Definition~\ref{def:cglp-proper-run}). The \temph{outcome} of a proper maGLP run is the tuple of per-agent outcomes.
\end{definition}

To relate maGLP to deduction, consider the \temph{lifted system} $L$: the cGLP transition system whose resolvent is the union of all agents' local resolvents, whose initial goal includes a \texttt{network} goal with channels paired to each agent's network channels, and whose program $M'$ is $M$ augmented with the GLP definition of \texttt{network}.

\begin{restatable}{proposition}{propMaGLPDeduction}
\label{prop:maglp-deduction}
Every proper maGLP run over $P$ and $M$ is simulated by a proper cGLP run of $L$ whose outcome is a logical consequence of $M'/?$.
\end{restatable}

Every maGLP run is safe, so correctness (Definition~\ref{def:ts}) reduces to liveness.

\begin{definition}[maGLP Correct Run]
\label{def:maglp-fair-run}
A maGLP run is \temph{correct} iff for every equivalence class $[t] \in T/{\sim}$, whenever $[t]$ is enabled some $t' \in [t]$ is eventually taken.
\end{definition}

\mypara{Implementation}
The implementation-ready deterministic variants of GLP and maGLP (the latter deterministic at the agent, not system, level), along with formal correctness proofs that they implement their respective specifications, are presented in a companion paper~\cite{shapiro2026implementing}.

\section{The Grassroots Social Graph}
\label{sec:ma-social-graph}

The grassroots social graph is the foundation upon which all other grassroots platforms are built. Nodes represent cryptographically self-identified agents; edges represent authenticated bidirectional channels; connected components arise spontaneously through befriending.

Each agent processes messages from user and network input streams and maintains an outputs list with one typed entry per destination --- \verb|user_output|, \verb|net_output|, and a \verb|friend_output| per friend --- on which the library routers \verb|send_user|, \verb|send_net|, and \verb|send_friend| dispatch. The program supports three protocols: (1) \emph{cold-call befriending}, where agents with no prior shared variables establish friendship by exchanging a response variable through the network; (2) \emph{friend-mediated introduction}, where a mutual friend creates a channel pair and sends each half to the respective parties, establishing a direct connection; and (3) \emph{text messaging} between established friends via named output streams.

The core of the program is the procedure \verb|agent/4|; the key clauses follow. Module declarations (\verb|exported|/\verb|imported|, \verb|#|-qualified calls) are elided from quoted clauses.

\mypara{Channels}
A \emph{channel} is an authenticated bidirectional communication structure \verb|ch(In, Out)|, where \verb|In| is a stream of incoming messages and \verb|Out| is a stream of outgoing messages. Two agents share a channel by holding opposite views of the same underlying stream pair: what one writes on its \verb|Out|, the other reads from its \verb|In|. The procedure \verb|new_channel/2| creates such a pair:
\begin{verbatim}
new_channel(ch(Xs?, Ys), ch(Ys?, Xs)).
\end{verbatim}
A single call \verb|new_channel(PQCh, QPCh)| binds \verb|PQCh| to \verb|ch(Xs?, Ys)| and \verb|QPCh| to \verb|ch(Ys?, Xs)|: the reader \verb|Xs?| of one channel is paired with the writer \verb|Xs| of the other (and likewise for \verb|Ys|/\verb|Ys?|), so each end's outgoing stream is the other end's incoming stream. Sending and receiving on a channel are:
\begin{verbatim}
send(X, ch(In, [X?|Out?]), ch(In?, Out)).
receive(X?, ch([X|In], Out?), ch(In?, Out)).
\end{verbatim}
Channels are used in friend-mediated introduction (below) to establish a direct connection between two parties without a cold-call. They are also used by the cold-call protocol: when the recipient accepts, it returns a freshly-created setup-channel half as its response; the friend channel itself is conveyed over the setup channel by the shared commit procedure \verb|befriend_commit|, completing the bidirectional link.

\mypara{Cold-call befriending}
The user's \verb|connect(Target)| request triggers a cold-call: the agent sends \verb|intro(Id, Resp)| on the network and injects the paired reader \verb|Resp?| into its own user-input stream, so the eventual response surfaces locally as a normal user-side message:
\begin{verbatim}
agent(Id, [msg('_user', Id1, connect(Target))|UserIn], NetIn, Outs) :-
    Id? =?= Id1?, ground(Target?) |
    send_net(msg(Target?, intro(Id?, Resp)), Outs?, Outs1),
    inject_msg(Resp?, Target?, Id?, UserIn?, UserIn1),
    agent(Id?, UserIn1?, NetIn?, Outs1?).
\end{verbatim}
On the recipient side, an incoming \verb|intro(From, Resp?)| captures the response writer \verb|Resp|, forwarded to the user in a befriend prompt; the user's decision will assign it, transparently transported back to the originator:
\begin{verbatim}
agent(Id, UserIn, [msg(Id1, intro(From, Resp?))|NetIn], Outs) :-
    Id? =?= Id1? |
    send_user(msg(agent, '_user', befriend(From?, Resp)), Outs?, Outs1),
    agent(Id?, UserIn?, NetIn?, Outs1?).
\end{verbatim}
When the user makes a decision (accept or reject), the agent processes the response:
\begin{verbatim}
agent(Id, [msg('_user', Id1, decision(Dec, From, response(Resp?)))|UserIn],
      NetIn, Outs) :-
    Id? =?= Id1? |
    bind_response(Id?, Dec?, From?, Resp, Outs?, Outs1, NetIn?, NetIn1),
    agent(Id?, UserIn?, NetIn1?, Outs1?).
\end{verbatim}
The helper \verb|bind_response| handles both cases: on \verb|yes|, it creates a fresh setup-channel pair via \verb|new_channel|, assigns \verb|accept(RetCh?)| to \verb|Resp| (transported back to the initiator by the Communicate transaction) and commits the friendship over the local half via the shared \verb|befriend_commit|, which conveys a fresh friend channel over the setup channel, registers its \verb|friend_output| entry, and merges its incoming end into \verb|NetIn|; on \verb|no|, it informs the user the request was rejected.  Both ends run the same idempotent commit.  Its definition appears in \appref{appendix:social-graph-walkthrough}{\arxivref}.

When Alice executes this protocol to befriend Bob, her agent sends \verb|msg(bob, intro(alice, Resp))| on the \verb|'_net'| output stream. The Cold-call transaction (Definition~\ref{definition:maGLP}) transfers this message to Bob's network input stream, adding the writer \verb|Resp| to Bob's resolvent. When Bob accepts, assigning \verb|Resp| to \verb|accept(RetCh?)| for a fresh setup channel, the Communicate transaction transfers this assignment back to Alice; \verb|befriend_commit| then installs the friend channel at both ends.

\mypara{Friend-mediated introduction}
Friend-mediated introduction creates a fresh setup-channel pair via \verb|new_channel/2| and sends one half to each of the introduced parties; on mutual acceptance a direct connection arises without a cold-call:
\begin{verbatim}
agent(Id, [msg('_user', Id1, introduce(P, Q))|UserIn], NetIn, Outs) :-
    Id? =?= Id1?, ground(P?), ground(Q?), ~(P? =?= Q?), new_channel(PQCh, QPCh) |
    send_friend(P?, msg(Id?, P?, intro(Q?, QPCh?)), Outs?, Outs1),
    send_friend(Q?, msg(Id?, Q?, intro(P?, PQCh?)), Outs1?, Outs2),
    agent(Id?, UserIn?, NetIn?, Outs2?).
\end{verbatim}
When Bob introduces Alice to Charlie, the setup-channel pair he creates contains readers that are transferred to Alice and Charlie via Communicate; each party accepts by acknowledging on the setup channel, and on mutual acknowledgement both commit the friendship via \verb|befriend_commit| --- a direct connection without any Cold-call.

\mypara{Text messaging}
Once two agents are friends, messages flow directly along their shared channel via the Communicate transaction, without cold-calls.  The user's \verb|send(Target, Text)| request is dispatched by \verb|send_friend| on the \verb|friend_output| entry registered under \verb|Target| in \verb|Outs|:
\begin{verbatim}
agent(Id, [msg('_user', Id1, send(Target, Text))|UserIn], NetIn, Outs) :-
    Id? =?= Id1?, ground(Target?) |
    send_friend(Target?, msg(Id?, Target?, text(Text?)), Outs?, Outs1),
    agent(Id?, UserIn?, NetIn?, Outs1?).
\end{verbatim}
On the other side, an incoming \verb|text(Text)| from \verb|From| is forwarded to the user as \verb|received(From, Text)|:
\begin{verbatim}
agent(Id, UserIn, [msg(From, Id1, text(Text))|NetIn], Outs) :-
    Id? =?= Id1? |
    send_user(msg(agent, '_user', received(From?, Text?)), Outs?, Outs1),
    agent(Id?, UserIn?, NetIn?, Outs1?).
\end{verbatim}

\mypara{Boot and deployment}
Each agent runs on a separate isolate (later --- separate smartphone). A boot clause reduces the universal initial state \verb|agent(ch(_?,_),ch(_?,_))| (Definition~\ref{definition:maGLP}) to the four-argument form used in the clauses above. The complete tested program --- the type vocabulary, helper procedures, the UI mediator, the actors, the plays, and the boot variants --- is the directory \verb|programs/social/graph| of the public GLP repository (\url{https://github.com/EShapiro2/GLP}); a comprehensive Alice/Bob/Charlie scenario appears in \appref{appendix:social-graph-walkthrough}{\arxivref}.

The social graph specified above, run as a maGLP program, is itself a grassroots platform: it inherits the grassroots property of maGLP (Corollary~\ref{corollary:social-graph-grassroots}).

\section{Multiagent GLP is Grassroots}\label{sec:grassroots}

This section establishes that maGLP is grassroots. We define the notion of grassroots following~\cite{lewis2026volitional}; formal definitions appear in \appref{appendix:grassroots-defs}{\arxivref}.

Informally, a protocol is \emph{grassroots} if two disjoint groups of agents can each operate independently---their interleaved correct runs are correct runs of the combined system---yet the combined system offers genuinely new behaviours that neither group could produce on its own. The first requirement is \emph{obliviousness}; the second is \emph{interactivity}.

The Cold-call transaction (Definition~\ref{definition:maGLP}) is the fundamental interactive transaction of maGLP. It allows an agent $q$ in one group to connect to an agent $p$ in another, establishing shared variables that span both groups---a behaviour that no interleaving of independent runs of the two groups could produce.

\begin{restatable}{theorem}{thmMaGLPGrassroots}
\label{theorem:maGLP-grassroots}
The maGLP protocol is grassroots.
\end{restatable}

The grassroots property of maGLP extends to applications built on top of it, provided they use the cold-call mechanism.

\begin{definition}[GLP Application]
A \temph{GLP application} is a GLP program $M$ together with the maGLP infrastructure. An application \temph{uses cold-calls} if agents can execute the Cold-call transaction to establish communication with previously-disconnected agents.
\end{definition}

\begin{restatable}{proposition}{propAppGrassroots}
\label{prop:app-grassroots}
Any GLP application that uses cold-calls is grassroots.
\end{restatable}

\begin{restatable}{corollary}{corSocialGraphGrassroots}
\label{corollary:social-graph-grassroots}
The GLP implementation of the grassroots social graph (Section~\ref{sec:ma-social-graph}) is grassroots.
\end{restatable}

\section{Conclusion}\label{section:conclusion}

While concurrent logic programming and its powerful programming techniques have been known for four decades~\cite{shapiro1989family}, adoption has been hampered by their inaccessibility to the average programmer. Concurrent logic programming requires a substantial shift from sequential procedural or functional thinking to dataflow programming via partial bindings, with concurrent operational semantics in which goals \emph{suspend} on unbound readers, \emph{commit} to clauses under guards, and \emph{communicate} through paired logical variables rather than explicit messages; reasoning about safety, fairness, deadlock-freedom, and progress in this setting is correspondingly demanding. With AI as a hyper-programmer this limitation is removed. Our experience to date is that AI is highly effective and productive at programming from mathematical specifications, particularly when the target is a typed high-level language such as GLP, where types constrain the space of legal programs and provide a checkable specification at the human-AI interface~\cite{shapiro2026types}. Moreover, our preliminary experience suggests that the abstract nature of concurrent logic programming renders it a more powerful and productive language than mainstream languages for collaborative human-AI abstract specifications-based program development.

Constitutional governance of grassroots communities and federations builds on Constitutional Consensus~\cite{keidar2025constitutional,lewis2025morpheus} and digital social contracts~\cite{cardelli2020digital,shapiro2022foundations}, both realisable in GLP. A textbook on GLP and programming in it is in preparation. The GLP runtime and example programs are open-source at \url{https://github.com/EShapiro2/GLP}.

\bibliographystyle{eptcs}
\bibliography{bib}

\ifappendix
\appendix

\section{Logic Programs}\label{appendix:lp}

We recall standard Logic Programs (LP) notions of syntax, most-general unifier (mgu), and semantics via goal reduction.

\begin{definition}[Logic Programs Syntax]
\label{def:lp-syntax}
We employ standard LP notions. Let $\calV$ denote the set of \temph{variables} (identifiers beginning with uppercase). A \temph{term} is a variable, a constant (numbers, strings, or the empty list \verb|[]|), or a compound term $f(T_1,\ldots,T_n)$ with functor $f$ and subterms $T_i$. Let $\calT$ denote the set of all terms. We use standard list notation: \verb=[X|Xs]= for a list cell, \verb|[X1,...,Xn]| for finite lists. A term is \temph{ground} if it contains no variables.

A \temph{unit goal} is a compound term or a string, also commonly referred to as an \temph{atom}. A \temph{goal} is a multiset of unit goals; the empty goal is written \verb|true|. A \temph{clause} $A$~\verb|:-|~$B$ has head $A$ (a unit goal) and body $B$ (a goal); a \temph{unit clause} has empty body. A \temph{logic program} is a finite set of clauses; clauses for the same predicate form a \temph{procedure}. Let $\calG(P)$ denote the set of goals over the vocabulary of the program $P$.
\end{definition}

A \emph{substitution} $\sigma$ is an idempotent function $\sigma: \calV \to \calT$, a mapping from variables to terms applied to a fixed point. By convention, $\sigma(x)=x\sigma$. Let $\Sigma$ denote the set of all substitutions. We assume standard notions of instance, ground, renaming, renaming apart, unifier, and most-general unifier (mgu).

\begin{definition}[LP Goal/Clause Reduction]
\label{def:lp-reduction}
Given an LP unit goal $A$ and clause $C$, with $H$ \verb|:-| $B$ being the result of renaming $C$ apart from $A$, the \temph{LP reduction} of $A$ with $C$ \temph{succeeds with} $(B,\sigma)$ if $A$ and $H$ have an mgu $\sigma$.
\end{definition}

\begin{definition}[Logic Programs Transition System]
\label{def:lp-ts}
A transition system $LP(P) = (C, c_0, T)$ is a \temph{Logic Programs transition system} for a logic program $P$ and initial goal $G_0 \in \mathcal{G}(P)$, if $C=\mathcal{G}(P)\times \Sigma$, $c_0=(G_0,\emptyset)$, and $T$ is the set of all transitions $(G,\sigma) \rightarrow (G',\sigma')$ such that for some unit goal $A \in G$ and clause $C \in P$ the LP reduction of $A$ with $C$ succeeds with $(B,\hat\sigma)$, $G' = (G \setminus \{A\} \cup B)\hat\sigma$, and $\sigma'=\sigma\circ\hat\sigma$.
\end{definition}

LP has two forms of nondeterminism: the choice of $A \in G$, called \emph{and-nondeterminism}, and the choice of $C \in P$, called \emph{or-nondeterminism}, and as such are closely-related to Alternating Turing Machines~\cite{shapiro1984alternation}.

\begin{definition}[Proper and Successful Run, Outcome]
\label{def:proper-run}
A run $\rho: (G_0,\sigma_0) \rightarrow \cdots \rightarrow (G_n, \sigma_n)$ of $LP(P)$ is \temph{proper} if for any $1\le i< n$, a variable that occurs in $G_{i+1}$ but not in $G_i$ also does not occur in any $G_j$, $j<i$. If proper, the \temph{outcome} of $\rho$ is $(G_0$ \verb|:-| $G_n)\sigma_n$. Such a run is \temph{successful} if $G_n=\emptyset$.
\end{definition}
The following proposition justifies the computation-as-deduction view of LP~\cite{kowalski1974predicate},
calling a proper LP run a \emph{derivation} and a complete proper run ending in the empty goal a \emph{successful derivation}.

\begin{restatable}[LP Computation is Deduction]{proposition}{propLPDeduction}
\label{prop:lp-deduction}
The outcome $(G_0$ \texttt{:-} $G_n)\sigma$ of a proper run of $LP(P)$ is a logical consequence of $P$.
\end{restatable}

The $LP(P)$ transition system allows defining several denotational semantic notions for a program $P$: (1) the \emph{clause semantics} is the set of all outcomes of all proper runs with an initial most-general unit goal (arguments are distinct variables), closely related to the fully-abstract compositional semantics of LP~\cite{gaifman1989fully}; (2) the \emph{atom semantics} is the set of all outcomes of all successful derivations with an initial most-general unit goal; (3) the \emph{ground atom semantics} is the standard model-theoretic semantics, the set of ground instances of the atom semantics over the Herbrand universe of $P$~\cite{lloyd1987foundations}.

\section{Term Matching}\label{appendix:term-matching}

If two terms $T_1$ and $T_2$ that jointly satisfy SO are unifiable with an mgu $\sigma$, then $\sigma$ maps any variable in $T_1$ to a subterm of $T_2$ and vice versa. Hence, the SO invariant of GLP allows eschewing unification in favour of \emph{term matching} that performs joint term-tree traversal and collects variable assignments along the way.

\begin{definition}[Term Matching]
\label{def:term-matching}
Given two terms $T_1$ and $T_2$ that jointly satisfy SO, their \temph{term matching} proceeds via the joint traversal of the term-trees of $T_1$ and $T_2$, consulting the following table at each pair of joint vertices, where $X_1, X_2$ denote writers, $X_1?, X_2?$ denote readers, and $f/n$ denotes a non-variable term, a constant when $n=0$ and a compound term when $n>0$:
\begin{center}
\begin{tabular}{l|lll}
$T_1 \backslash T_2$ & Writer $X_2$ & Reader $X_2?$ & Term $f_2/n_2$ \\
\hline
Writer $X_1$ & fail & $X_1 := X_2?$ & $X_1 := T_2$ \\
Reader $X_1?$ & $X_2 := X_1?$ & fail & suspend on $X_1?$\\
Term $f_1/n_1$ & $X_2 := T_1$ & fail & fail if $f_1 \ne f_2$ or $n_1 \ne n_2$\\
\end{tabular}
\end{center}
The writer mgu is the union of all writer assignments if no \emph{fail} was encountered and the suspension set is empty.
\end{definition}

\begin{remark}
In an actual implementation, assuming $T_1$ is a goal term and $T_2$ a head term, the case of $X_1?$ and $T_2$ would add $X_1?$ to the set of readers the goal would suspend upon.
\end{remark}

\section{Grassroots Definitions}\label{appendix:grassroots-defs}

This appendix contains the full definitions for protocols and the grassroots property referenced in Section~\ref{sec:grassroots}, following~\cite{lewis2026volitional}.

\subsection{Protocols and the Grassroots Property}

A protocol is a family of multiagent transition systems, one for each finite set of agents $P \subset \Pi$, sharing an underlying set of local states with a designated initial state.

\begin{definition}[Local-States Function]\label{definition:local-states-function}
A \temph{local-states function} $S: 2^\Pi \to 2^\calS$ maps every finite set of agents $P \subset \Pi$ to a set of local states $S(P) \subset \calS$ that includes $s_0$ and satisfies $P \subset P' \implies S(P) \subset S(P')$.
\end{definition}

Given a local-states function $S$, we use $C(P) := S(P)^P$ for configurations over $P$ and $c_0(P) := \{s_0\}^P$ for the initial configuration.

\begin{definition}[Protocol]\label{definition:protocol}
A \temph{protocol} $\calF$ over a local-states function $S$ is a family of multiagent transition systems with exactly one transition system $\calF(P) = (C(P), c_0(P), T(P), {\sim(P)})$ for every finite $P \subset \Pi$.
\end{definition}

A \emph{correct} run of a transition system is safe and live (Definition~\ref{def:ts}); the runs here are safe by construction, so correctness coincides with liveness.

Informally, in a grassroots protocol two disjoint groups of agents can each operate independently---their interleaved correct runs are correct runs of the combined system---yet the combined system offers genuinely new behaviours that neither group could produce on its own. To define this formally, we first define the interleaving of runs of two disjoint groups.

\begin{definition}[Interleaving]\label{def:interleaving}
Let $P, P' \subset \Pi$ be disjoint nonempty sets of agents, $r = c_0, c_1, \ldots$ a run of $\calF(P)$, and $r' = d_0, d_1, \ldots$ a run of $\calF(P')$. An \temph{interleaving} of $r$ and $r'$ is a sequence $e_0, e_1, \ldots$ of configurations in $C(P \cup P')$ for which there exist non-decreasing sequences of indices $(i_k)_{k \geq 0}$ and $(j_k)_{k \geq 0}$ with $i_0 = j_0 = 0$ such that for every $k \geq 0$:
\begin{enumerate}
\item $(e_k)_p = (c_{i_k})_p$ for every $p \in P$,
\item $(e_k)_q = (d_{j_k})_q$ for every $q \in P'$,
\item if $e_{k+1}$ exists, then exactly one of:
  (a) $i_{k+1} = i_k + 1$ and $j_{k+1} = j_k$ (a $P$-step), or
  (b) $i_{k+1} = i_k$ and $j_{k+1} = j_k + 1$ (a $P'$-step).
\end{enumerate}
\end{definition}

Note that an interleaving is well-defined: since $P \subset P \cup P'$, the local-states function ensures $S(P) \subset S(P \cup P')$, and similarly for $P'$, so each $e_k$ is a valid configuration in $C(P \cup P')$. Also, $e_0 = c_0(P \cup P')$, since $(e_0)_p = (c_0)_p = s_0$ for all $p \in P$ and $(e_0)_q = (d_0)_q = s_0$ for all $q \in P'$.

\begin{definition}[Oblivious, Interactive, Grassroots]\label{definition:grassroots}
A protocol $\calF$ is:
\begin{enumerate}
\item \temph{oblivious} if for every disjoint nonempty $P, P' \subset \Pi$,
    every interleaving of a correct run of $\calF(P)$ and a correct run of $\calF(P')$ is a correct run of $\calF(P \cup P')$.
\item \temph{interactive} if for every disjoint nonempty $P, P' \subset \Pi$, there exists a correct run $\hat{r}$ of $\calF(P \cup P')$ such that for every correct run $r$ of $\calF(P)$, every correct run $r'$ of $\calF(P')$, and every interleaving $e$ of $r$ and $r'$, $\hat{r} \neq e$.
\item \temph{grassroots} if it is oblivious and interactive.
\end{enumerate}
\end{definition}

Being oblivious means that two disjoint groups of agents, each running the protocol independently and correctly, do not interfere with each other: any interleaving of their independent correct runs is a correct run of the combined system. Being interactive means that two disjoint groups, when brought together, can do something genuinely new: there exists a correct run of the combined system that could not arise from the two groups operating independently. In an interleaving, each step changes the local states of agents in only one group; therefore, any transaction whose active participants span both groups yields a step that cannot occur in any interleaving. These definitions are the machine-layer specialisation, without guards, of the volitional grassroots framework of~\cite{lewis2026volitional}.

\subsection{Transactions-Based Protocols}

\begin{definition}[Transactions Over a Local-States Function]\label{definition:transactions-lsf}
Let $S$ be a local-states function. A set of transactions $R$ is \temph{over $S$} if every transaction $t \in R$ is a transition over participants $Q$ and local states $S(Q')$ for some $Q \subseteq Q' \subset \Pi$. Given such $R$ and $P \subset \Pi$:
$$R(P) := \{t \in R : t \text{ has participants } Q \subseteq P \text{ and is over } S(Q') \text{ for some } Q' \subseteq P\}$$
\end{definition}

\begin{definition}[Transactions-Based Protocol]\label{definition:tb-protocol}
Let $S$ be a local-states function and $R$ a set of transactions over $S$ with equivalence $\sim$. The \temph{protocol $\calF$ over $R$, $S$, and $\sim$} is defined by $\calF(P) := (C(P), c_0(P), R(P){\uparrow}P, {\sim})$ for each $P \subset \Pi$.
\end{definition}

Since liveness applies to all equivalence classes, any class $[t]$ in $R(P \cup P')/{\sim}$ whose transactions have participants spanning both $P$ and $P'$ could obstruct obliviousness if enabled in an interleaving. The following proposition identifies the condition under which this does not occur.

\begin{restatable}[Obliviousness Condition]{proposition}{propOblivious}
\label{proposition:oblivious}
A transactions-based protocol is oblivious provided that for every disjoint nonempty $P, P' \subset \Pi$, no equivalence class whose transactions have participants spanning both $P$ and $P'$ is ever enabled in any interleaving of correct runs of $\calF(P)$ and $\calF(P')$.
\end{restatable}

\begin{restatable}[Transactions-Based Grassroots]{theorem}{thmInteractiveGrassroots}
\label{theorem:interactive-grassroots}
A transactions-based protocol that satisfies the condition of Proposition~\ref{proposition:oblivious} and is interactive is grassroots.
\end{restatable}

\section{Guards and System Predicates}\label{appendix:guards-system}

Guards and system predicates extend GLP programs with access to the GLP runtime state, operating system and hardware capabilities.

\mypara{Predefined types}
The predefined types, available in every GLP program, are \verb|Stream(X)|, \verb|OpenStream(X)|, \verb|DiffList(X)|, \verb|Channel(In, Out)|, \verb|Constant|, and the arithmetic-expression type \verb|Exp|; their definitions are in \verb|programs/self.glp| of the public GLP repository. \verb|Number|, \verb|Integer|, \verb|Real|, and \verb|String| are primitive types built into the runtime.

\mypara{Guard predicates}
Guards provide read-only access to the runtime state of GLP computation. A guard appears after the clause head, separated by \verb=|=, and must be satisfied for the clause to be selected. The ``Ground'' column indicates whether success of the guard implies that the argument is ground; such guards permit multiple occurrences of the reader in the clause body (Remark~\ref{rem:guards-srsw}). Built-in guards can be negated: \verb|~G| succeeds if \verb|G| fails and vice versa, while suspension is unchanged; negation applies only to atomic built-in guards, not to defined guards or conjunctions.

\begin{center}
\begin{tabular}{lll}
\textbf{Guard} & \textbf{Signature} & \textbf{Ground} \\
\hline
\verb|integer| & \verb|procedure integer(Integer?).| & yes \\
\verb|number| & \verb|procedure number(Number?).| & yes \\
\verb|string| & \verb|procedure string(String?).| & yes \\
\verb|constant| & \verb|procedure constant(Constant?).| & yes \\
\verb|compound| & \verb|procedure compound(_?).| & no \\
\verb|list| & \verb|procedure list(_?).| & no \\
\verb|ground| & \verb|procedure ground(_?).| & yes \\
\verb|known| & \verb|procedure known(_?).| & no \\
\verb|unknown| & \verb|procedure unknown(_?).| & no \\
\verb|no_readers| & \verb|procedure no_readers(_?).| & no \\
\verb|module| & \verb|procedure module(_?).| & yes \\
\verb|@<| & \verb|procedure @<(Constant?, Constant?).| & yes \\
\verb|is_mutual_ref| & \verb|procedure is_mutual_ref(_?).| & no \\
\verb|otherwise| & (no arguments) & --- \\
\verb|=?=| & \verb|procedure =?=(_?, _?).| & yes (both) \\
\end{tabular}
\end{center}

\verb|otherwise| succeeds if all previous clauses for this procedure failed (or suspended).

\verb|=?=| succeeds if both arguments are ground and equal. For example, \verb|f(a,X?) =?= f(b,Z?)| fails (not suspends) because the ground subterms \verb|a| and \verb|b| already differ. Note that the implementation checks terms left-to-right and does not guarantee early failure detection; \verb|f(X?,a) =?= f(Y?,b)| may suspend rather than fail.

\mypara{Arithmetic comparison guards}
Arithmetic comparison guards evaluate their arguments as arithmetic expressions and compare the results. Success implies both arguments are ground.

\begin{center}
\begin{tabular}{ll}
\textbf{Guard} & \textbf{Signature} \\
\hline
\verb|<| & \verb|procedure <(Exp?, Exp?).| \\
\verb|>| & \verb|procedure >(Exp?, Exp?).| \\
\verb|=<| & \verb|procedure =<(Exp?, Exp?).| \\
\verb|>=| & \verb|procedure >=(Exp?, Exp?).| \\
\verb|=:=| & \verb|procedure =:=(Exp?, Exp?).| \\
\verb|=\=| & \verb|procedure =\=(Exp?, Exp?).| \\
\end{tabular}
\end{center}

\mypara{Time guards}
Time guards provide access to the system clock for scheduling and synchronization:

\begin{center}
\begin{tabular}{ll}
\textbf{Guard} & \textbf{Signature} \\
\hline
\verb|wait| & \verb|procedure wait(Number?).| \\
\verb|wait_until| & \verb|procedure wait_until(Number?).| \\
\end{tabular}
\end{center}
\noindent\verb|wait(D)| suspends for \verb|D| milliseconds then succeeds. \verb|wait_until(T)| suspends until the current time is at or past timestamp~\verb|T| (Unix milliseconds), then succeeds. Neither guard can be negated.

\mypara{Monotonicity and implications}
\verb|ground/1|, \verb|no_readers/1|, and \verb|known/1| are monotonic. \verb|ground(X)| implies both \verb|no_readers(X)| and \verb|known(X)|, which do not imply each other: \verb|no_readers(X)| succeeds but \verb|known(X)| fails for an unassigned writer; \verb|no_readers(f(X?))| suspends but \verb|known(f(X?))| succeeds.

\mypara{Defined guard predicates}
To support abstract data types and cleaner code organization, GLP provides for user-defined guards via unit clauses. A unit clause \verb|p(T1,...,Tn).| defines a guard predicate; the call \verb|p(S1,...,Sn)| in guard position is unfolded to the term matching of \verb|T1| with \verb|S1|, ..., \verb|Tn| with \verb|Sn|. For example, unification is defined by the unit clause \verb|X? = X.|, so the guard \verb|A = B| unfolds to matching both \verb|A| and \verb|B| against the same variable \verb|X|.

The following defined guard predicates are shipped with the runtime:

\begin{center}
\begin{tabular}{ll}
\textbf{Predicate} & \textbf{Unit clause} \\
\hline
\verb|=| & \verb|X? = X.| \\
\verb|new_channel| & \verb|new_channel(ch(Xs?, Ys), ch(Ys?, Xs)).| \\
\verb|send| & \verb|send(X, ch(In, [X?|Out?]), ch(In?, Out)).| \\
\verb|receive| & \verb|receive(X?, ch([X|In], Out?), ch(In?, Out)).| \\
\verb|dl_append| & \verb|dl_append(A\B?, B\C?, A?\C).| \\
\verb|dl_to_list| & \verb|dl_to_list(L\[], L?).| \\
\end{tabular}
\end{center}

\mypara{Body kernel predicates}
\temph{Body kernel predicates} are runtime-implemented primitives that execute in the clause body with two-valued semantics: they either succeed or abort. Body kernels are not directly accessible to user programs. They are invoked only by \emph{system predicates}---predicates defined by GLP clauses that are shipped with the runtime and have privileged access to the body kernel registry. The system predicate's own guards ensure that a body kernel's preconditions are met before invocation.

Body kernels follow a naming convention of quoted atoms beginning with an underscore (e.g., \verb|'_add'|), preventing collisions with user-defined predicates. The following table lists all body kernels:

\begin{center}
{\small
\begin{tabular}{lll}
\textbf{Category} & \textbf{Kernel} & \textbf{Operation} \\
\hline
Arithmetic & \verb|'_add'(Number?, Number?, Number)| & Addition \\
 & \verb|'_sub'(Number?, Number?, Number)| & Subtraction \\
 & \verb|'_mul'(Number?, Number?, Number)| & Multiplication \\
 & \verb|'_div'(Number?, Number?, Number)| & Division (real result) \\
 & \verb|'_idiv'(Integer?, Integer?, Integer)| & Integer division \\
 & \verb|'_mod'(Integer?, Integer?, Integer)| & Modulo \\
 & \verb|'_neg'(Number?, Number)| & Unary negation \\
 & \verb|'_abs'(Number?, Number)| & Absolute value \\
\hline
Math & \verb|'_sqrt'(Number?, Number)| & Square root \\
 & \verb|'_sin'(Number?, Number)| & Sine \\
 & \verb|'_cos'(Number?, Number)| & Cosine \\
 & \verb|'_tan'(Number?, Number)| & Tangent \\
 & \verb|'_asin'(Number?, Number)| & Arc sine \\
 & \verb|'_acos'(Number?, Number)| & Arc cosine \\
 & \verb|'_atan'(Number?, Number)| & Arc tangent \\
 & \verb|'_exp'(Number?, Number)| & Exponential ($e^x$) \\
 & \verb|'_ln'(Number?, Number)| & Natural logarithm \\
 & \verb|'_log10'(Number?, Number)| & Base-10 logarithm \\
 & \verb|'_pow'(Number?, Number?, Number)| & Power ($x^y$) \\
\hline
Conversion & \verb|'_integer'(Number?, Integer)| & Convert to integer \\
 & \verb|'_real'(Number?, Number)| & Convert to float \\
 & \verb|'_round'(Number?, Integer)| & Round to nearest integer \\
 & \verb|'_floor'(Number?, Integer)| & Floor \\
 & \verb|'_ceil'(Number?, Integer)| & Ceiling \\
\hline
Structure & \verb|'_list_to_tuple'(List?, _)| & List to compound term \\
 & \verb|'_tuple_to_list'(_?, List)| & Compound term to list \\
\hline
Time & \verb|'_now'(Integer)| & Current Unix timestamp (ms) \\
\hline
I/O & \verb|'_output'(_?)| & Send term to output callback \\
\hline
MWM & \verb|'_allocate_mutual_reference'(_, _)| & Create mutual reference for O(1) append \\
 & \verb|'_stream_append'(_?, _?, _)| & Append value via mutual reference \\
 & \verb|'_close_mutual_reference'(_?)| & Close stream (bind tail to \verb|[]|) \\
\hline
Network & \verb|'_send'(_?, _?, _?)| & Send madGLP message to remote agent \\
\hline
Modules & \verb|'_activate'(Module?, _)| & Dispatch goal against module's \verb|_select/1| \\
\end{tabular}
}
\end{center}

\mypara{System predicates}
System predicates are predicates defined by GLP clauses whose bodies may invoke body kernel predicates. They provide safe, user-accessible interfaces to runtime services:

\begin{center}
\begin{tabular}{ll}
\textbf{Predicate} & \textbf{Description} \\
\hline
\verb|=..| & Term composition and decomposition (bidirectional) \\
\verb|now/1| & Current Unix timestamp (milliseconds) \\
\verb|:=/2| & Arithmetic evaluation and assignment \\
\end{tabular}
\end{center}

\mypara{Term composition and decomposition (\texttt{=..})}
The \verb|=..| predicate is bidirectional: when the right-hand side is a list, it composes a compound term; when the left-hand side is a compound term, it decomposes it into a list. The reader/writer positions of the arguments determine the direction. For example, \verb|T =.. [foo, a, b]?| constructs \verb|T = foo(a, b)|, and \verb|foo(a, b) =.. L| produces \verb|L = [foo, a, b]|.

\mypara{Clock access (\texttt{now/1})}
Binds the argument to the current Unix timestamp in milliseconds since the epoch.

\mypara{Arithmetic evaluation and assignment (\texttt{:=})}
Arithmetic assignment is defined recursively: each clause pattern-matches on an arithmetic operator, guards verify that operands are of the required type, and the body invokes the corresponding body kernel. An \verb|otherwise| clause handles the case where operands are themselves expressions, evaluating them recursively before retrying.

The complete definitions of \verb|:=|, \verb|now|, and \verb|=..| are in \verb|programs/self.glp| of the public GLP repository.

\section{Deferred Proofs}\label{appendix:proofs}

This appendix contains deferred proofs from Sections~\ref{sec:glp}, \ref{sec:maglp}, and~\ref{sec:grassroots}.

\subsection{cGLP Proofs}
\label{app:glp-proofs}

\propGLPDeduction*

\begin{proof}
The $/\!?$ operator replaces every reader $X?$ by its paired writer $X$, transforming GLP terms into LP terms. We show that the cGLP run $\rho$ corresponds to an LP run $\rho/?$ of $LP(P/?)$.

Consider a cGLP transition $(G, \sigma) \rightarrow (G', \sigma')$:
\begin{itemize}
    \item \emph{Reduce transition}: Goal $A$ reduces with clause $C$ via writer mgu $\hat\sigma$. Applying $/\!?$, the clause $C/?$ is an LP clause, and $A/?$ unifies with the head $H/?$ via the mgu $\hat\sigma/?$ (since writers map to writers). This is a valid LP reduction.
    \item \emph{Communicate transition}: A reader $X? \in G$ is replaced by the value $T$ assigned to its paired writer. Under $/\!?$, both $X?$ and $X$ map to $X$, so this transition becomes the identity---the variable $X$ is already assigned $T$ in the LP view.
\end{itemize}

Thus each cGLP transition corresponds to zero or one LP transitions, and the cGLP run $\rho$ projects to an LP run $\rho/?$ of $LP(P/?)$. By standard LP soundness, the outcome of $\rho/?$ is a logical consequence of $P/?$.
\end{proof}

\propSOPreservation*

\begin{proof}
By induction on the length of the run. The base case is immediate: $G_0$ satisfies SO by assumption.

For the inductive step, assume $G$ satisfies SO and consider a transition $(G, \sigma) \rightarrow (G', \sigma')$:
\begin{itemize}
    \item \emph{Reduce transition}: Goal $A \in G$ reduces with clause $C = (H \mathrel{\mbox{\texttt{:-}}} B)$ via writer mgu $\hat\sigma$, yielding $G' = (G \setminus \{A\} \cup B)\hat\sigma$. Since $C$ satisfies SRSW, it satisfies SO. Since $C$ is renamed apart from $G$, the variables in $B$ are fresh. The writer mgu $\hat\sigma$ maps writers in $A$ to subterms of $H$ and vice versa; by SO of both $G$ and $C$, each variable is assigned at most once. Applying $\hat\sigma$ to $(G \setminus \{A\} \cup B)$ replaces each variable by a term containing fresh variables (from $B$) or ground subterms. Since no variable in $G \setminus \{A\}$ occurs in $A$ (by SO of $G$), and no variable in $B$ occurs in $G$ (by renaming apart), $G'$ satisfies SO.

    \item \emph{Communicate transition}: $G' = G\hat\sigma?$ where $\hat\sigma? = \{X? := T\}$. Since $G$ satisfies SO, $X?$ occurs at most once in $G$. Replacing this single occurrence by $T$ (which satisfies SO by Definition~\ref{def:writers-assignment}) preserves SO, provided $T$ shares no variables with the rest of $G$. By the proper run condition (Definition~\ref{def:cglp-proper-run}), variables in $T$ are fresh, so $G'$ satisfies SO.
\end{itemize}
\end{proof}

\propMonotonicity*

\begin{proof}
Suppose goal $A$ can reduce with clause $C$ at step $i$, meaning the writer mgu of $A$ and the head $H$ of (a renaming of) $C$ succeeds. Consider what can change between steps $i$ and $j > i$:
\begin{itemize}
    \item \emph{Reduce transitions on other goals}: These do not affect $A$ directly. By SO, no other goal shares a writer with $A$, so no other reduction can assign a writer in $A$.

    \item \emph{Communicate transitions}: These instantiate readers, not writers. A Communicate transition $X? := T$ may instantiate a reader $X? \in A$, yielding $A' = A\{X? := T\}$. We show $A'$ can still reduce with $C$:

    The original writer mgu succeeded, meaning at position $p$ where $X?$ occurred in $A$, $H$ had a writer $Y$ at position $p$, yielding assignment $Y := X?$; a reader or a non-variable term of $H$ at position $p$ would have caused failure or suspension, respectively, contradicting the success of the writer mgu.

    After the Communicate transition, $A'$ has $T$ at position $p$. The clause $C$ (renamed apart from $A'$) has a fresh writer $Y'$ at position $p$. The writer mgu now yields $Y' := T$, which succeeds.

    \item \emph{Reduce transition on $A$}: If $A$ itself is reduced at some step $k$ with $i < k \le j$, then an instance of $A$ has been reduced, satisfying the proposition.
\end{itemize}

Thus, if $A$ has not been reduced by step $j$, the (possibly instantiated) goal $A'$ at step $j$ can still reduce with a fresh renaming of $C$.
\end{proof}

\subsection{maGLP Proofs}
\label{app:maglp-proofs}

\propMaGLPSOPreservation*

\begin{proof}
Reduce $p$ preserves SO at agent $p$ by the same argument as cGLP SO Preservation (Proposition~\ref{prop:so-preservation}), restricted to agent $p$'s resolvent. Communicate $p$ to $q$ removes $\{X? := T\}$ from $\sigma_p$ (no change to $G_p$) and applies $\{X? := T\}$ to $G_q$. The substitution preserves SO in $G_q$ provided variables in $T$ are fresh with respect to $G_q$; this holds by the proper run condition, since $X$ was a fresh writer when assigned. For Cold-call $p$ to $q$, the message \verb|msg|$(q,X)$ is transferred from $p$'s network output stream to $q$'s network input stream. The writers in $X$ are no longer reachable from $p$'s resolvent after advancing (the resolvent has iterated past the consumed stream element), so they occur only in $q$'s resolvent. SO is preserved.
\end{proof}

\propMaGLPDeduction*

\begin{proof}
Let $r = c^0 \rightarrow c^1 \rightarrow \cdots$ be a proper maGLP run over $P$ and $M$, with $c^k_p = (G^k_p, \sigma^k_p)$ for $p \in P$. We construct a proper cGLP run $\hat r$ of $L$ such that, after simulating the $k$-th maGLP step, the goal of $\hat r$ contains $\bigcup_{p \in P} G^k_p$ and its substitution contains $\bigcup_{p \in P} \sigma^k_p$.

The initial goal of $L$ is $\bigcup_{p \in P} G^0_p$ together with the \verb|network| goal. It satisfies SO: within each agent's initial goal every variable occurs once, the agents' initial goals are variable-disjoint, and each agent's network-channel variables are paired one-for-one with those of the \verb|network| goal. Hence by SO Preservation (Proposition~\ref{prop:so-preservation}) every goal of $\hat r$ satisfies SO. We simulate each maGLP transaction.

\emph{Reduce $p$.} The step is a cGLP Reduce of some $A \in G^k_p$ with a clause $C \in M \subseteq M'$. Since $G^k_p$ is part of the goal of $\hat r$ and $C \in M'$, the identical reduction is a cGLP Reduce step of $L$, updating the goal and the substitution exactly as in maGLP and leaving the other agents' goals and the \verb|network| goal untouched.

\emph{Communicate $p$ to $q$.} Here $\{X? := T\} \in \sigma^k_p$ and $X?$ occurs in $G^k_q$. Both lie in the substitution and the goal of $\hat r$, so the cGLP Communicate applying $\{X? := T\}$ is a step of $L$; as $X?$ occurs once in the goal, it instantiates exactly the occurrence in $q$'s goal, as in maGLP. The maGLP step removes $\{X? := T\}$ from $\sigma^k_p$ whereas the cGLP rule retains it, but $X?$ then occurs nowhere, so neither any subsequent goal nor the outcome is affected. When $p = q$ the transaction is unary and is itself a cGLP Communicate step of $L$.

\emph{Cold-call $p$ to $q$.} The network output stream of $c^k_p$ has a new message \verb|msg|$(q, X)$, and the step adds \verb|msg|$(q, X)$ to the network input stream of $c^k_q$. In $L$ the \verb|network| goal consumes \verb|msg|$(q, X)$ from $p$'s network output stream and places \verb|msg|$(q, X)$ on $q$'s network input stream, routing by the network-switch technique, generalised to $P$. This is a bounded sequence of cGLP steps of $L$ --- a Communicate delivering \verb|msg|$(q, X)$ to the \verb|network| goal, the Reduce of the routing clause, and a Communicate placing \verb|msg|$(q, X)$ on $q$'s network input stream --- after which \verb|msg|$(q, X)$ occurs on that stream, as in maGLP.

Each maGLP step is simulated by cGLP transitions of $L$, so $\hat r$ is a run of $L$. Reduce introduces only variables renamed apart and Communicate introduces none, so $\hat r$ inherits properness from $r$. By cGLP Computation is Deduction (Proposition~\ref{prop:cglp-deduction}) applied to $L$ over $M'$, the outcome of $\hat r$ is a logical consequence of $M'/?$.
\end{proof}

\subsection{Grassroots Proofs}
\label{app:grassroots-proofs}

\propOblivious*

\begin{proof}
Let $\calF$ be a transactions-based protocol over transactions $R$, local-states function $S$, and equivalence $\sim$. Let $P, P' \subset \Pi$ be disjoint and nonempty, $r = c_0, c_1, \ldots$ a correct run of $\calF(P)$, $r' = d_0, d_1, \ldots$ a correct run of $\calF(P')$, and $e = e_0, e_1, \ldots$ an interleaving of $r$ and $r'$.

\emph{Safety.} We show that $e$ is a run of $\calF(P \cup P')$. We have $e_0 = c_0(P \cup P')$ as noted in the well-definedness remark following Definition~\ref{def:interleaving}. Consider a $P$-step $e_k \rightarrow e_{k+1}$. The corresponding transition $c_{i_k} \rightarrow c_{i_k+1}$ is in $T(P) = R(P){\uparrow}P$, so there exists a transaction $t \in R(P)$ with participants $Q \subseteq P$. Since $P \subseteq P \cup P'$, we have $R(P) \subseteq R(P \cup P')$, so $t \in R(P \cup P')$. In the step $e_k \rightarrow e_{k+1}$, the $Q$-agents change as specified by $t$ and all other agents---both $P \setminus Q$ and $P'$---remain stationary. By the definition of closure, $e_k \rightarrow e_{k+1} \in t{\uparrow}(P \cup P') \subseteq T(P \cup P')$. The case of a $P'$-step is symmetric.

\emph{Liveness.} We show that $e$ is live. Consider an equivalence class $[t]$ in $R(P \cup P')/{\sim}$.
\begin{enumerate}
\item If all transactions in $[t]$ have participants $Q \subseteq P$: whether $[t]$ is enabled in $e_k$ depends only on the states of agents in $Q \subseteq P$, which in the interleaving are identical to those in the $P$-run at the corresponding index. Since $r$ is correct and hence live with respect to $[t]$, the interleaving is live with respect to $[t]$.
\item If all transactions in $[t]$ have participants $Q \subseteq P'$: symmetric.
\item If $[t]$ contains transactions with participants spanning both $P$ and $P'$: by assumption, $[t]$ is never enabled in the interleaving, so liveness holds vacuously.
\end{enumerate}
\end{proof}

\thmInteractiveGrassroots*

\begin{proof}
By Proposition~\ref{proposition:oblivious}, the protocol is oblivious. If it is also interactive, it is grassroots by Definition~\ref{definition:grassroots}.
\end{proof}

\thmMaGLPGrassroots*

\begin{proof}
By Theorem~\ref{theorem:interactive-grassroots}, it suffices to show that maGLP is a transactions-based protocol satisfying the obliviousness condition and that it is interactive.

\emph{maGLP is transactions-based}: By Definition~\ref{definition:maGLP}, maGLP is defined via transactions (Reduce, Communicate, Cold-call) over a local-states function with a common initial state and the transaction equivalence defined for maGLP in Section~\ref{sec:maglp-def}.

\emph{Obliviousness condition}: Reduce is unary, so only Communicate and Cold-call may have participants spanning both $P$ and $P'$. A Cold-call$(p, q)$ with $p \in P$, $q \in P'$ requires $p$ to have a network output message addressed to $q$; in a run of $\calF(P)$, agent $p$ never produces messages for agents outside $P$, so no such Cold-call is ever enabled in an interleaving. A Communicate$(p, q)$ with $p \in P$, $q \in P'$ requires $p$ and $q$ to share a variable pair, which can only be established by a prior Cold-call with participants in both groups; since no such Cold-call is ever enabled, no such Communicate is ever enabled either. The condition of Proposition~\ref{proposition:oblivious} is satisfied.

\emph{Interactivity}: Let $P, P' \subset \Pi$ be disjoint and nonempty. Let $\hat{r}$ be a correct run of maGLP$(P \cup P')$ in which some $p \in P$ performs a Cold-call to some $q \in P'$, establishing shared reader/writer pairs between $p$ and $q$. The Cold-call step changes the local states of both $p$ and $q$, and hence is neither a $P$-step nor a $P'$-step. The run $\hat{r}$ is correct: any Communicate class enabled by the resulting shared variables can be taken, and any Reduce class thereby enabled can be taken in turn. Therefore $\hat{r}$ is not an interleaving of any runs of $P$ and $P'$ separately, and maGLP is interactive.
\end{proof}

\propAppGrassroots*

\begin{proof}
A GLP application using cold calls is a restriction of maGLP to a specific program $M$. The transactions-based structure, the obliviousness condition, and the Cold-call-based interactivity argument of the proof of Theorem~\ref{theorem:maGLP-grassroots} all carry over. Hence by Theorem~\ref{theorem:interactive-grassroots}, the application is grassroots.
\end{proof}

\corSocialGraphGrassroots*

\begin{proof}
The social graph uses cold calls for initial contact between disconnected agents. By Proposition~\ref{prop:app-grassroots}, it is grassroots.
\end{proof}

\section{Social Graph Walkthrough}\label{appendix:social-graph-walkthrough}

This appendix presents the detailed walkthrough of the grassroots social graph program introduced in Section~\ref{sec:ma-social-graph}.

\subsection{Types and Channels}

The protocol type vocabulary --- \verb|Response|, \verb|Decision|, \verb|NetColdCall|, \verb|FriendContent|, \verb|IntroContent|, \verb|OutputEntry|, \verb|OutputsList|, and the rest --- is defined in \verb|self.glp| of the repository directory (Section~\ref{sec:ma-social-graph}).

A \verb|Channel| is a pair of streams for bidirectional communication. The \verb|new_channel| guard, shown in Section~\ref{sec:ma-social-graph}, creates complementary channel endpoints: each reads from its first stream and writes to its second, and each end's outgoing stream is the other end's incoming stream.

\subsection{Agent Structure}

Each agent processes messages from two separate input streams---user and network---and maintains a typed outputs list (Section~\ref{sec:ma-social-graph}). The four arguments of \verb|agent/4| are: agent identity, user input stream, network input stream, and the outputs list. The outputs list initially contains two entries---\verb|user_output(...)| and \verb|net_output(...)|---providing streams to the user interface and the network, respectively. As the agent befriends others, \verb|friend_output| entries are added.

\subsection{Cold-Call Befriending Protocol}

The cold-call protocol enables agents to establish friendship without prior shared variables. Its initiating, receiving, and decision clauses appear in Section~\ref{sec:ma-social-graph}; the remaining clause, handling the response to a sent cold-call, is in \verb|agent.glp| of the repository directory.

The protocol works as follows: (1) Alice sends \verb|connect(bob)| to her agent via the \verb|'_user'| stream; (2) her agent sends a 2-argument \verb|msg(bob, intro(alice, Resp))| via the \verb|'_net'| output, including a fresh response variable \verb|Resp|; (3) Bob receives \verb|msg(bob, intro(alice, Resp))| on his network input and forwards the request to his user interface; (4) Bob decides \verb|yes| or \verb|no|; (5) Bob's agent creates a setup-channel pair and sends \verb|accept(RetCh?)| back via \verb|Resp|; (6) Alice's agent receives the response, and both agents commit the friendship via the shared \verb|befriend_commit|, which conveys a fresh friend channel over the setup channel and adds a \verb|friend_output| entry on each side.

\subsection{Channel Establishment}

When a cold-call is accepted, both agents establish symmetric channels; the procedures \verb|bind_response| and \verb|handle_response| are in \verb|agent.glp| of the repository directory.

The accepting agent creates a setup-channel pair via \verb|new_channel(RetCh, LocalCh)|, sends \verb|RetCh| back to the initiator, and commits over \verb|LocalCh|. The shared library procedure \verb|befriend_commit| creates the fresh friend channel, conveys it as \verb|friend_channel(...)| over the setup channel, adds the \verb|friend_output| entry, and merges the friend's incoming stream into the agent's network input; it is idempotent, so both ends run the same commit.

\subsection{Text Messaging}

Once agents are friends, they can exchange text messages; the sending and receiving clauses appear in Section~\ref{sec:ma-social-graph}.

\subsection{Friend-Mediated Introduction}

Once agents are friends, they can introduce each other to third parties. The introducing clause, which creates a fresh channel pair and sends each half to the respective parties, appears in Section~\ref{sec:ma-social-graph}; the receiving-side clauses are in \verb|agent.glp| of the repository directory.

When Bob types \verb|introduce(alice, charlie)|, he creates a setup-channel pair via \verb|new_channel(PQCh, QPCh)|; Alice and Charlie each receive one half. Accepting sends \verb|ack| on the setup channel; \verb|intro_await_peer| waits for the peer's \verb|ack| (or \verb|nack|) and injects the result, and on mutual acknowledgement each side commits via \verb|befriend_commit|. When both accept, they become direct friends without Bob's further involvement.

Note the distinction between cold-call messages and friend-mediated messages: cold-call uses a 2-argument \verb|msg(Target, Content)| sent via the \verb|'_net'| output, while friend-mediated introduction uses a 3-argument \verb|msg(From, To, intro(Other, Ch))| sent directly to a friend via their named output.

\subsection{Network Switch Simulation}

In deployment, agents communicate through a physical network. In simulation, a \verb|network3| process routes messages between three agents; it is listed in full in \verb|boot.glp| of the repository directory (Section~\ref{sec:ma-social-graph}).

The network switch routes both 2-argument cold-call messages (where \verb|msg(Target, Content)| is addressed by target name) and 3-argument friend-to-friend messages (where \verb|msg(From, To, Content)| carries both sender and receiver identities). In deployment, the network switch is replaced by maGLP's Cold-call transaction (Section~\ref{sec:maglp}).

\subsection{The Scenario and Actors}

The complete scenario demonstrates all three protocols:
\begin{enumerate}
\item Alice cold-calls Bob (Bob accepts) --- Alice and Bob become friends
\item Alice sends Bob: ``Hi Bob, this is Alice''
\item Bob cold-calls Charlie (Charlie accepts) --- Bob and Charlie become friends
\item Charlie sends Bob: ``Hi Bob, this is Charlie''
\item Bob introduces Alice to Charlie (both accept) --- Alice and Charlie become direct friends
\item Alice sends Charlie: ``Hi Charlie, this is Alice''
\item Charlie responds: ``Hi Alice, this is Charlie''
\end{enumerate}

Each agent is driven by an \emph{actor}---a GLP procedure that implements a state machine, reacting to messages from the UI mediator and producing commands. The actor communicates in ground terms only---the same protocol a human uses in the interactive UI. The actor's state is encoded in procedure names, with transitions via recursive calls. The complete program --- helper procedures, the UI mediator, the actors, the plays that tie everything together, and the boot variants --- is the repository directory \verb|programs/social/graph| (Section~\ref{sec:ma-social-graph}).

\fi

\end{document}